\documentclass[aps,prd,superscriptaddress,twoside,twocolumn,nofootinbib,showpacs,
floatfix]{revtex4-2}
\usepackage{amsmath,amssymb}
\usepackage{graphicx,bm}
\usepackage{slashed,color}
\usepackage{ulem,slashed,tensor,braket}
\usepackage[colorlinks=true,pdfstartview=FitV,bookmarks=true,bookmarksnumbered=true,
breaklinks]{hyperref}
\usepackage{cancel}
\hypersetup{linkcolor=red, citecolor=blue, urlcolor=blue}

\usepackage[dvipsnames]{xcolor} 
\begin{document}
\title{Rescattering effects in near-threshold $J/\psi$ photoproduction}
\author{S. Sakinah}
\affiliation{Department of Physics, Kyungpook National University, Daegu 41566,
South Korea}
\author{Sang-Ho Kim}
\email{shkimphy@gmail.com}
\affiliation{Department of Physics and Origin of Matter and Evolution of Galaxies
(OMEG) Institute, Soongsil University, Seoul 06978, South Korea}
\author{H. M. Choi}
\affiliation{Department of Physics, Kyungpook National University, Daegu 41566,
South Korea}

\date{\today}

\begin{abstract}

We investigate near-threshold $J/\psi$ photoproduction off the nucleon, focusing on
hadronic rescattering effects induced by open-charm meson-baryon intermediate
states.
Beyond the conventional Pomeron-exchange mechanism, the $\bar D^0\Lambda_c^+$ and
$\bar D^{*0} \Lambda_c^+$ channels are incorporated within an effective Lagrangian
framework.
The relevant production amplitudes are evaluated at tree level from $t$-, $s$-, and
$u$-channel diagrams in a gauge-invariant manner.
The resulting total and $t$-dependent differential cross sections are compared with
recent near-threshold data from the GlueX, $J/\psi$-007, and CLAS experiments at
Jefferson Lab.
We find that the open-charm rescattering contributions significantly improve the
description of the data, particularly at large momentum transfer, and naturally
generate cusp-like structures near the $\bar D^0 \Lambda_c^+$ and $\bar D^{*0}
\Lambda_c^+$ thresholds in the GlueX data.
We further present predictions for the associated open-charm processes $\gamma p
\to \bar D^{(*)0} \Lambda_c^+$, whose cross sections are estimated to be of the order
of 5 nb.

\end{abstract}

\maketitle
\section{Introduction}

Near-threshold photoproduction of the $J/\psi$ meson off the nucleon has attracted
considerable interest as a sensitive probe of the nucleon’s gluonic structure
and the QCD dynamics of the charmonium–nucleon interaction.
Early measurements were performed at SLAC~\cite{Camerini:1975cy} and
Fermilab~\cite{Binkley:1981kv,E687:1993hlm}, and later extended to much higher
energies at HERA, where the H1~\cite{H1:1996kyo,H1:2000kis} and ZEUS~\cite{
ZEUS:2002wfj} Collaborations measured exclusive $\gamma p \to J/\psi p$ cross
sections up to $W \approx$ 300 GeV.
In this high-energy regime, the data are well described by soft- and hard-Pomeron
exchanges within the Donnachie–Landshoff (DL) model~\cite{Donnachie:1984xq,
Donnachie:1985iz,Donnachie:1987pu,Donnachie:1998gm,Donnachie:1999qv}, as well as by
the soft-dipole Pomeron approach~\cite{Fiore:1998jx,Martynov:2002ez}.

More recently, Jefferson Lab (JLab) has carried out high-precision measurements near
threshold (4.0 $\leqslant W \leqslant$ 4.8 GeV) through the GlueX~\cite{
GlueX:2019mkq,GlueX:2023pev}, $J/\psi$-007~\cite{Duran:2022xag,007:2026dow}, and
CLAS~\cite{CLAS:2026lls} experiments.
These data provide new constraints on the reaction mechanisms underlying $J/\psi$
photoproduction~\cite{Lee:2022ymp}.
A variety of theoretical descriptions have been proposed, including the two-gluon
exchange model~\cite{Zeng:2020coc}, perturbative QCD approaches based on the
nucleon's generalized parton distributions (GPDs)~\cite{Guo:2021ibg}, and scalar
($0^{++}$)- and tensor ($2^{++}$)-glueball exchanges within holographic QCD
frameworks~\cite{Hatta:2018ina,Hatta:2019lxo,Mamo:2019mka,Mamo:2021tzd}.
Complementary to these gluon-based approaches, Ref.~\cite{Sakinah:2024cza} suggests
that the $J/\psi N$ final-state interaction (FSI) may play an important role very
close to threshold ($W \leqslant$ 4.2 GeV), based on a phenomenological
quark-nucleon potential combined with Pomeron exchange.
In Refs.~\cite{Tang:2024pky,Tang:2025qqe}, the $\gamma \to c \bar c + \mathbb{P} \to
J/\psi$ transition amplitude is formulated in terms of dressed quark degrees of
freedom, following the formalism of Ref.~\cite{Pichowsky:1996tn}.

Beyond these partonic descriptions, Ref.~\cite{Kim:2025oyo} considered tree-level
$t$-channel exchanges involving both light mesons and charmonium states.
Although light mesons and charmonium exhanges are generally expected to be
suppressed by the Okubo-Zweig-Iizuka (OZI) rule~\cite{Okubo:1963fa,Okubo:1963fa2,
Okubo:1963fa3} and by the heavy-quark mass, respectively, the light-meson
contribution can nevertheless be non-negligible.
In addition, the $J/\psi N$ FSI mediated by gluon exchange can further enhance the
cross section very close to threshold.

Open-charm meson–baryon intermediate states have been studied extensively~\cite{
Wu:2010jy,Liu:2019tjn,Fernandez-Ramirez:2019koa,Du:2019pij}.
They are known to play an essential role in the formation of heavy pentaquark
candidates~\cite{Wang:2015jsa,Kubarovsky:2015aaa,Karliner:2015voa,HillerBlin:2016odx,
Wang:2019krd,Wu:2019adv,Winney:2019edt,Strakovsky:2023kqu} and may also influence
near-threshold $J/\psi$ photoproduction.
The closest open-charm channels, $\bar D^0 \Lambda_c^+$ and $\bar D^{*0} \Lambda_c^+$,
lie approximately 116 MeV and 258 MeV above the $J/\psi$ photoproduction threshold,
respectively~\cite{PDG:2024cfk}.
Whether cusp structures associated with these channels exist in this energy region
remains under debate.
For example, cusp-like structures near these thresholds have been reported by the
GlueX Collaboration~\cite{GlueX:2019mkq,GlueX:2023pev} and supported by theoretical
studies~\cite{Du:2020bqj,JPAC:2023qgg}.
Meanwhile, very recent measurements of the differential cross sections through the
dimuon decay channel ($J/\psi \to \mu^+ \mu^-$) have been reported by the
$J/\psi$-007 Collaboration~\cite{007:2026dow}.
However, the corresponding integrated cross sections show no clear evidence for
open-charm contributions.
Measurements of the total cross section by the CLAS Collaboration likewise suggest
only a small contribution from open-charm channels, with a more gradual energy
dependence~\cite{CLAS:2026lls}.

In parallel, the LHCb Collaboration reported several hidden-charm pentaquark
states in the $J/\psi p$ invariant-mass spectrum of $\Lambda_b \to J/\psi p K^-$
decays, including one narrow state below the $\bar D \Sigma_c$ threshold, $P_{c \bar c}
(4312)$, and two narrow states below the $\bar D^* \Sigma_c$ threshold, $P_{c \bar c}
(4440)$ and $P_{c \bar c} (4457)$~\cite{LHCb:2015yax,LHCb:2019kea}.
More recently, Ref.~\cite{LHCb:2021chn} reported a new state, $P_{c \bar c} (4330)$, in
$B_s^0 \to J/\psi p \bar p$ decays, while the previously reported broad state
$P_{c \bar c} (4380)$ below the $\bar D \Sigma_c^*$ threshold was not observed.
These observations have motivated coupled-channel studies that explore the role of
open-charm channels in the formation of heavy pentaquark states~\cite{
Clymton:2024fbf,Zhang:2024dkm,Wu:2024xwy}.

We investigate the $\gamma p \to J/\psi p$ reaction by incorporating hadronic
rescattering contributions~\cite{Sato:1996gk,Oh:2003aw} beyond the conventional
Pomeron-exchange mechanism~\cite{Donnachie:1984xq,Donnachie:1985iz,Donnachie:1987pu,
Donnachie:1998gm,Donnachie:1999qv}.
Specifically, we include the $\bar D^0 \Lambda_c^+$ and $\bar D^{*0} \Lambda_c^+$
open-charm meson–baryon intermediate states and evaluate all relevant $t$-, $s$-, and
$u$-channel diagrams at tree level within an effective Lagrangian framework, while
preserving gauge invariance for each subamplitude.
We show that the inclusion of these rescattering contributions improves the
description of recent near-threshold JLab data and provides a natural explanation for
the cusp-like structures seen in the data.
We also present predictions for the open-charm production processes, $\gamma p \to
\bar D^{(*)0} \Lambda_c^+$.

The present work is organized as follows.
In {\color{red}Sec.~\ref{Sec:II}}, we outline the general formalism for the Pomeron-exchange and
rescattering amplitudes in the $\gamma p \to J/\psi p$ reaction, and describe in
detail how the rescattering diagrams are constructed.
In Sec.~\ref{Sec:III}, we present the results for the total and $t$-dependent
differential cross sections and compare them with the JLab data. 
Predictions for open-charm production are also given.
Finally, we summarize our findings and present our conclusions in the last section.

\section{Theoretical Framework}
\label{Sec:II}
In this section, we present the theoretical framework for the process
\begin{align}
\gamma(q) + p(p_i) \to J/\psi(k) + p(p_f).
\label{eq:reaction}
\end{align}
We first consider Pomeron exchange as the background mechanism within the DL
model~\cite{Donnachie:1984xq,Donnachie:1985iz,Donnachie:1987pu,Donnachie:1998gm,
Donnachie:1999qv}.
We then incorporate rescattering contributions involving the intermediate $\bar D^0
\Lambda_c^+$ and $\bar D^{*0}\Lambda_c^+$ states in a gauge-invariant manner.
The general formalism for the $\bar D^0\Lambda_c^+$ and $\bar D^{*0}\Lambda_c^+$
rescattering contributions is presented in Secs.~\ref{Sec:II-1}
and~\ref{Sec:II-2}, respectively

\subsection{Pomeron-exchange amplitude}
\label{Sec:II-1}

\begin{figure}[ht]
\centering
\includegraphics[width=3.5cm]{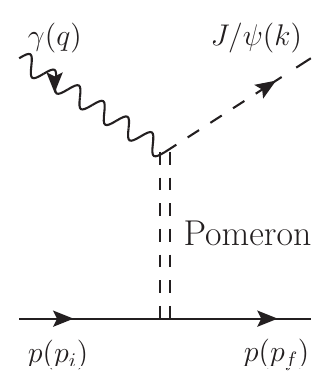}
\caption{Two-gluon exchange mechanism of $\gamma p \to J/\psi p$.}
\label{FIG01}
\end{figure}

The two-gluon exchange mechanism shown in Fig.~\ref{FIG01} is effectively described
by the DL model~\cite{Donnachie:1984xq,Donnachie:1985iz,Donnachie:1987pu,
Donnachie:1998gm,Donnachie:1999qv}.
In this picture, the incoming photon first fluctuates into a $q\bar q$ pair, which
subsequently interacts with the nucleon via Pomeron exchange before forming the final
$J/\psi$ meson.
The quark-Pomeron vertex is constructed by analogy with the Pomeron-photon coupling,
treating the Pomeron as a $C = +1$ isoscalar photon.

The Pomeron-exchange amplitude can be written as
\begin{align}
T_{\mathbb{P}} (q,k) &= i\frac{12e}{f_V} \beta_{q_V} F_{V}(t) \beta_{u/d} F_N(t)
\cr
& \times
\bar u_N \varepsilon_\nu^* [ \slashed{q}\,g^{\mu\nu} - \gamma^\mu q^\nu ]
\epsilon_\mu u_N
G_{\mathbb{P}}(s,t),
\label{eq:Amp:Pom}
\end{align}
where $\epsilon_\mu$ and $\varepsilon_\nu$ are the polarization vectors of the
incoming photon and outgoing $J/\psi$ meson, respectively.
Here, $u_N$ denotes the nucleon spinor, $e$ is the elementary electric charge, and
$f_V$ is the vector-meson decay constant defined in the vector meson dominance (VMD)
convention; for the $J/\psi$, we use $f_{J/\psi} = 11.18$~\cite{Kim:2025oyo}.
All baryon Dirac spinors are normalized as $\bar u_B u_B = 1$.
The Mandelstam variables are defined by $s=W^2=(q+p_i)^2$, $t=(q-k)^2$, and
$u=(q-p_f)^2$.
The coupling constants $\beta_{u/d}$ = 2.07 and $\beta_c$ = 0.323 GeV$^{-1}$ are
chosen to reproduce the total cross-section data for diffractive photoproduction of
vector mesons at high photon energies.

The form factor for the Pomeron-vector-meson vertex and the isoscalar electromagnetic
(EM) form factor of the nucleon are defined as~\cite{Donnachie:1999qv,Titov:2003bk}
\begin{align}
F_V (t) &= \frac{2 \mu_0^2}{(1-t/M_V^2)(2\mu_0^2+M_V^2-t)},
\cr
F_N(t) &= \frac{4M^2_N-2.8t}{(4M^2_N-t)(1-t/0.71)^2} ,
\label{eq:FF:PomExch}
\end{align}
respectively, where $\mu_0^2$ = 1.1 GeV$^2$, and $M_V$ and $M_N$ denote the
vector-meson and nucleon masses, respectively.

The Regge propagator for the Pomeron is of the form~\cite{Donnachie:1998gm}
\begin{align}
G_{\mathbb{P}} (s,t) =
\left( \frac{s}{s_0} \right)^{\alpha_{\mathbb P}(t)-1}
\exp \left\{ -\frac{i\pi}{2} [\alpha_{\mathbb P}(t)-1] \right\}
\label{eq:Prop:SPom},
\end{align}
where the Pomeron trajectory is taken as $\alpha_{\mathbb{P}}(t)$ = 1.25 +
$\alpha_{\mathbb P}' t$, with $\alpha_{\mathbb P}'$ = $1/s_0$ = 0.25 GeV$^{-2}$.

\subsection{Rescattering amplitudes}
\label{Sec:II-2}

\begin{figure}[!]
\centering
\includegraphics[width=7.5cm]{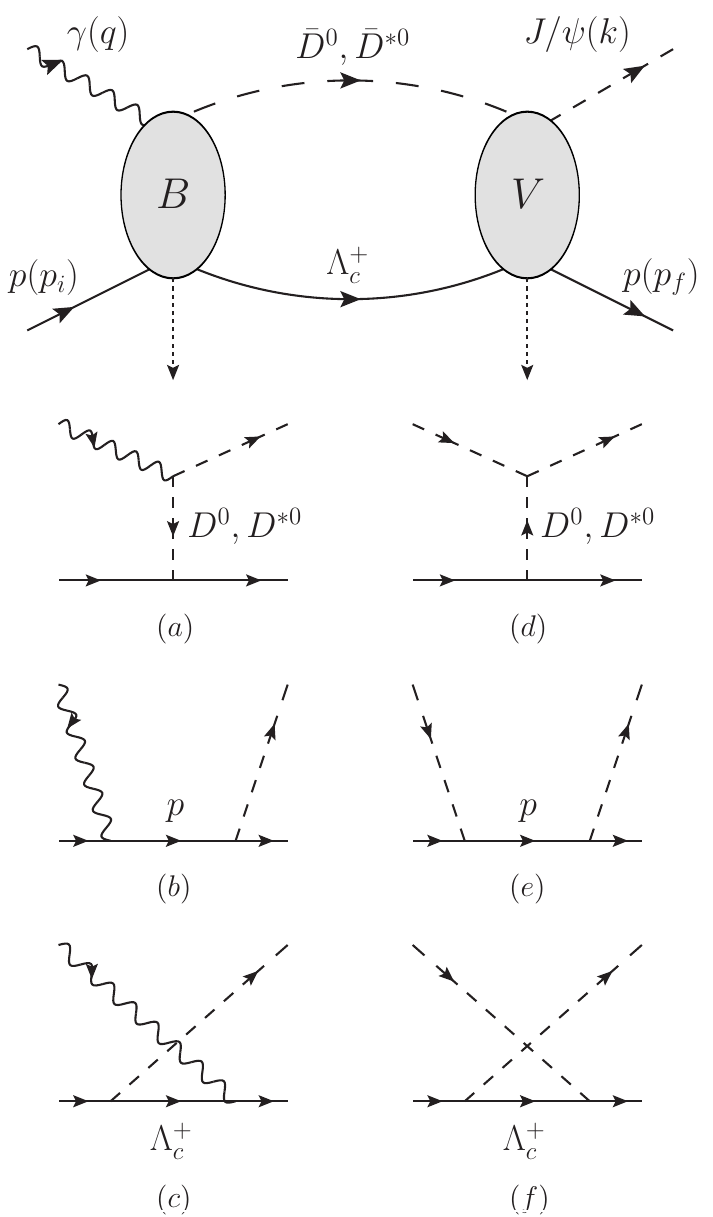}
\caption{Feynman diagrams for the $\gamma p \to \bar D^{(*)0} \Lambda_c^+ \to J/\psi p$
rescattering process.
Panels $(a)–(c)$ and $(d)–(f)$ show the $t$-, $s$-, and $u$-channel tree diagrams for
the production amplitude $B_{\gamma p \to MB}$ and the transition amplitude
$V_{MB \to J/\psi p}$, respectively.}
\label{FIG02}
\end{figure}

The meson--baryon ($MB$) rescattering amplitudeis given
by~\cite{Sato:1996gk,Oh:2003aw}
\begin{align}
T_{\mathrm{Res}}(q,k) &= \int d^3 \mathbf{q'}
B_{\gamma p \to MB}(q, q') \frac{1}{2 E_M ({\bf q'})} \frac{M_B}{E_B ({\bf q'})}
\cr
& \times
G_{MB} (\mathbf{q'}) V_{MB \to J/\psi p}(q',k).
\label{eq:ResAmpl-1}
\end{align}
Equation~(\ref{eq:ResAmpl-1}) represents the two-step process in which the initial
$\gamma p$ state first produces an intermediate meson--baryon state $MB$, which then
propagates and rescatters into the final $J/\psi p$ state.
Here, $B_{\gamma p \to MB}$ denotes the production amplitude for the intermediate
state, $G_{MB}$ the meson--baryon propagator, and $V_{MB \to J/\psi p}$ the
transition amplitude for the rescattering process $MB \to J/\psi p$.
The corresponding Feynman diagram is shown in Fig.~\ref{FIG02}.

The meson--baryon propagator in Eq.~(\ref{eq:ResAmpl-1}) is given by
\begin{align}
G_{MB} (\mathbf{q'}) =
\frac{1}{W - E_M({\bf q'}) - E_B({\bf q'}) + i \epsilon},
\end{align}
where $E_M = \sqrt{M_M^2+|\mathbf{q'}|^2} $ and $E_B = \sqrt{M_B^2+|\mathbf{q'}|^2}$
are, respectively, the energies of the intermediate meson and baryon with
three-momentum $\mathbf{q'}$, and $W$ is the total center-of-mass (c.m.) energy.
The invariant amplitudes $B_{\gamma p \to MB}$ and $V_{MB \to J/\psi p}$ 
are constructed from $t$-, $s$-, and $u$-channel tree diagrams, as shown in
Fig.~\ref{FIG02}($a$)–($c$) and Fig.~\ref{FIG02}($d$)–($f$), respectively.
In the present work, we consider the intermediate states $(M,B) =
(\bar D^0,\Lambda_c^+)$ and $(\bar D^{*0},\Lambda_c^+)$.

For simplicity, we write $B(q,q') \equiv B_{\gamma p \to MB}(q,q')$ and $V(q',k) \equiv
V_{MB \to J/\psi p}(q',k)$.
Eq.~(\ref{eq:ResAmpl-1}) can be rewritten as
\begin{align}
& T_{\mathrm{Res}}(q,k) =
\cr
& \mathcal{P} \int 
\frac{d^3 \mathbf{q'}}{2 E_M ({\bf q'})} \frac{M_B}{E_B ({\bf q'})}
\frac{B(q, q') V(q', k)}{W - E_M({\bf q'}) - E_B({\bf q'})}
\cr
& - i \int 
\frac{d \Omega_{q_0}}{2 E_M ({\bf q_0})} \frac{M_B}{E_B ({\bf q_0})}
\rho_{MB} ({\bf q_0}) B(q, q_0) V(q_0,k)
\cr
&\hspace{1cm} \times \theta(W - M_M - M_B),
\label{eq:ResAmpl-2}
\end{align}
where $\theta(x)$ is the step function, $d\Omega_{q_0}$ denotes the solid-angle element
of the on-shell momentum $\mathbf{q}_0$, and $\rho_{MB}$ is given by
\begin{align}
\rho_{MB} ({\bf q_0}) = \frac{\pi |{\bf q_0|}
E_M({\bf q_0}) E_B({\bf q_0})}{E_M({\bf q_0}) + E_B({\bf q_0})}.
\label{eq:rho}
\end{align}
From the on-shell condition $W = E_M({\bf q_0}) + E_B({\bf q_0})$, the magnitude of the
on-shell momentum, $|\bf q_0|$, is given by
\begin{align}
|{\bf q_0}| = \frac{\sqrt{{[W^2 - (M_M + M_B) ^2]}{[W^2 - (M_M - M_B)^2]}}}{2 W}.
\label{eq:3-momentum}
\end{align}
The first term in Eq.~(\ref{eq:ResAmpl-2}) corresponds to the principal-value part,
while the second term represents the on-shell absorptive contribution above the
$MB$ threshold.
Summation over the polarization and spin indices of the intermediate states is implied.

The invariant amplitudes $B_{\gamma p \to MB}$ and $V_{MB \to J/\psi p}$ are constructed from
the effective Lagrangians.
The EM interaction can be written as
\begin{align}
\mathcal{L}_{\gamma D D^*} &= g_{\gamma D D^*}^0 \epsilon^{\mu\nu\alpha\beta}
\partial_\mu A_\nu
( \partial_\alpha D^{*0}_\beta \bar D^0 + \partial_\alpha \bar D^{*0}_\beta D^0),
\cr
\mathcal{L}_{\gamma D^* D^*} &=
- i g_{\gamma D^* D^*}^0 F^{\mu\nu} \bar D^*_\mu D^*_\nu,
\cr
\mathcal{L}_{\gamma NN} &=
- e \bar N \left[ \gamma_\mu - \frac{\kappa_N}{2M_N} \sigma_{\mu\nu} \partial^\nu
\right] A^\mu N,
\cr
\mathcal{L}_{\gamma \Lambda_c \Lambda_c} &=
- e \bar \Lambda_c \left[\gamma_\mu -\frac{\kappa_{\Lambda_c}}{2M_{\Lambda_c}}
\sigma_{\mu\nu}\partial^\nu \right] A^\mu \Lambda_c,
\end{align}
where $A^\mu$, $D$, $D^{*\mu}$, $\Lambda_c^+$, and $N$ denote the photon, pseudoscalar
$D$, vector $D^*$, $\Lambda_c$, and nucleon fields, respectively.
Here the $D$, $D^*$, and $\Lambda_c$ correspond to the $D(1864,0^-)$, $D^*(2007,1^-)$,
and $\Lambda_c(2286,1/2^+)$ states, respectively.
The EM field-strength tensor is defined by $F^{\mu\nu} = \partial^\mu A^\nu -
\partial^\nu A^\mu$.
Because the exchanged $D^{(*)0}$ mesons are neutral [see Fig.~\ref{FIG02}(a)], only
the magnetic-type EM couplings contribute, through the
$\gamma D D^*$ and $\gamma D^* D^*$ interaction terms.

The relevant coupling constants are given by
\begin{align}
g_{\gamma D D^*}^0 &= 0.54\, \mathrm{GeV}^{-1},
\cr
g_{\gamma D^* D^*}^0 &= 0.64,
\cr
\kappa_p &= 1.79,
\cr
\kappa_{\Lambda_c} &= -0.025.
\end{align}
The coupling $g_{\gamma D D^*}^0$ is extracted from the branching ratio $\mathrm{Br}
({D^{*0} \to D^0 \gamma}) $ = 35.3\,\%~\cite{PDG:2024cfk},
through the relation
\begin{align}
\Gamma_{D^{*0} \to D^0 \gamma} =
\frac{q_\gamma^3}{12\pi} (g_{\gamma D D^*}^0)^2,
\label{eq:Wid:GMJP}
\end{align}
with $q_\gamma = (M_{D^{*0}}^2-M_{D^0}^2)/(2M_{D^{*0}})$.
The total decay width of the $D^{*0}$ meson, which has not yet been measured, is
estimated using isospin symmetry~\cite{Du:2020bqj}.
The quantities $\kappa_p$~\cite{PDG:2024cfk} and $\kappa_{\Lambda_c}$~\cite{
Fomin:2019wuw} denote the anomalous magnetic moments of the proton and the $\Lambda_c$
baryon, respectively.

The effective Lagrangians involving the $J/\psi$ meson are given by
\begin{align}
\mathcal{L}_{\psi D D} &=
- i g_{\psi D D} \psi^\mu ( \bar D \partial_\mu D - \partial_\mu \bar D D),
\cr
\mathcal{L}_{\psi D D^*} &=
- g_{\psi D D^*} \epsilon^{\mu\nu\alpha\beta} \partial_\mu \psi_\nu
( \partial_\alpha D^*_\beta \bar D +
\partial_\alpha \bar D^*_\beta D),
\cr
\mathcal{L}_{\psi D^* D^*} &=
i g_{\psi D^* D^*} \psi^\mu
( D^{*\nu} \partial_\nu \bar D^*_\mu - \partial_\nu D^*_\mu \bar D^{*\nu}
\cr
& - D^{*\nu} \overleftrightarrow{\partial}_\mu \bar D^*_\nu ),
\cr
\mathcal{L}_{\psi \Lambda_c \Lambda_c} &=
g_{\psi \Lambda_c \Lambda_c} \bar \Lambda_c \gamma_\mu \psi^\mu \Lambda_c,
\cr
\mathcal{L}_{\psi NN} &=
-g_{\psi NN} {\bar N}  \gamma_\mu \psi^\mu N,
\end{align}
where $\psi^\mu$ denotes the $J/\psi(3096,1^-)$ field.
The coupling constants are obtained from VMD~\cite{Oh:2007ej}, yielding
\begin{align}
& g_{\psi D D} = g_{\psi D^* D^*} = 7.44,
\cr
& g_{\psi D D^*} = 3.84\, \mathrm{GeV}^{-1},
\end{align}
while $g_{\psi \Lambda_c \Lambda_c}$ is determined using SU(4) flavor symmetry~\cite{
Liu:2001ce}:
\begin{align}
g_{\psi \Lambda_c \Lambda_c} = -1.4.
\end{align}
The $\psi pp$ coupling constant is extracted from the branching ratio
$\mathrm{Br} (J/\psi \to p \bar p) = 0.2120\,\%$ and the total decay width
$\Gamma_{J/\psi} = 92.6$ keV~\cite{PDG:2024cfk}.
Using these values, we obtain
\begin{align}
|g_{\psi NN}| = |g_{\psi pp}| = 2.18 \times 10^{-3},
\end{align}
through the relation
\begin{align}
\Gamma_{J/\psi \to p \bar p} =
\frac{2}{3\pi} \frac{q_N^3}{M_{J/\psi}^2} g_{\psi p p}^2,
\label{eq:Wid:JPNN}
\end{align}
with $q_N = \sqrt{M_{J/\psi}^2- 4M_N^2}/2$.

The effective Lagrangians for the meson--baryon interactions are given by
\begin{align}
\mathcal{L}_{D N \Lambda_c} &=
- i g_{D N \Lambda_c} D \bar \Lambda_c \gamma_5 N + \mathrm{H.c.},
\cr
\mathcal{L}_{D^* N \Lambda_c} &=
- g_{D^* N \Lambda_c} D^{* \mu} \bar \Lambda_c \gamma_\mu N + \mathrm{H.c.}.
\end{align}
The coupling constants are determined using SU(4) flavor symmetry~\cite{Liu:2001ce}:
\begin{align}
g_{D N \Lambda_c} = -13.5,\,\,\,
g_{D^* N \Lambda_c} = -5.6.
\end{align}
These values are close to those obtained from light-cone sum rule (LCSR)
approach~\cite{Khodjamirian:2011jp,Khodjamirian:2011sp}:
$g_{D N \Lambda_c} = -10.7,\,g_{D^* N \Lambda_c} = -5.8$.

We now describe how the invariant amplitudes $B_{\gamma p \to MB}$ and $V_{MB \to J/\psi p}$
in Eq.~(\ref{eq:ResAmpl-1}) are constructed by considering the intermediate
$\bar D^0 \Lambda_c^+$ and $\bar D^{*0} \Lambda_c^+$ states.
We also describe how gauge invariance is preserved when hadronic form factors are
introduced.
The relevant four-momenta are defined as
\begin{align}
\gamma(q) + p(p_i) \to \bar D^{(*)0}(q') + \Lambda_c^+(p') \to J/\psi(k) + p(p_f).
\end{align}

\subsubsection{$\bar D^0 \Lambda_c^+$ intermediate state}

We first consider the invariant amplitude $B_{\gamma p \to \bar D^0 \Lambda_c^+}$.
For the $t$-channel diagram shown in Fig.~\ref{FIG02}(a), $D^0$ exchange is forbidden
by charge conservation, whereas the $D^*$-exchange amplitude is gauge invariant by
construction.
The $s$- and $u$-channel contributions in Figs.~\ref{FIG02}(b,\,c) are not
individually gauge invariant, but their sum is.
Accordingly, the $\gamma p \to \bar D^0 \Lambda_c^+$ invariant amplitude is
constructed in a gauge-invariant form as follows:
\begin{align}
B_{\gamma p \to \bar D^0 \Lambda_c^+} =
\mathcal{M}_{B,t}^{D^*} \, F_t^{D^*}
+ (\mathcal{M}_{B,s}^p + \mathcal{M}_{B,u}^{\Lambda_c}) F_{s,u}^C.
\label{B-Ampl-1}
\end{align}
Equation~(\ref{B-Ampl-1}) shows that the full amplitude is written as the sum of two
separately gauge-invariant parts: the $D^*$-exchange $t$-channel contribution and the
combined $s$- and $u$-channel contributions. Since the latter become gauge invariant
only through their sum, a common form factor $F_{s,u}^C$ is introduced for the
combined $(s+u)$ contribution in order to preserve gauge invariance after including
hadronic form factors.

The $t$-, $s$-, and $u$-channel amplitudes are given by
\begin{align}
\mathcal{M}^{D^*}_{B,t} &=
- \frac{g_{\gamma D D^*}^0 g_{D^* N \Lambda_c}}{q_t^2 - M_{D^*}^2}
\epsilon^{\mu\nu\alpha\beta}
\bar u_{\Lambda_c} \gamma_\mu q_{t \nu} q_\alpha \epsilon_\beta u_N,
\cr
\mathcal{M}^p_{B,s} &=
i \frac{e g_{D N \Lambda_c}}{q_s^2-M_N^2}
\bar u_{\Lambda_c} \gamma_5 (\slashed{q}_s + M_N)
\cr
& \times
\left[ \slashed{\epsilon} - \frac{\kappa_p}{4M_N}
(\slashed{\epsilon} \slashed{q} -\slashed{q} \slashed{\epsilon})
\right] u_N,
\cr
\mathcal{M}^{\Lambda_c}_{B,u} &=
i \frac{e g_{D N \Lambda_c}}{q_u^2-M_{\Lambda_c}^2} \bar u_{\Lambda_c}
\left[ \slashed{\epsilon} - \frac{\kappa_{\Lambda_c}}{4M_{\Lambda_c}}
(\slashed{\epsilon} \slashed{q} -\slashed{q} \slashed{\epsilon}) \right]
\cr
& \times
(\slashed{q}_u + M_{\Lambda_c}) \gamma_5 u_N,
\label{B-Ampl-2}
\end{align}
where $\epsilon_\mu$ is the polarization vector of the incoming photon,
and the exchanged four-momenta are defined by
\begin{align}
q_t = q - q',\,\,\, q_s = q + p_i,\,\,\, q_u = p_i - q'.
\label{FourMomenta-1}
\end{align}

Because of the finite size of hadrons, we introduce the hadronic form factors in
Eq.~(\ref{B-Ampl-1}):
\begin{align}
F_t^{D^{(*)}} &=
\left( \frac{\Lambda_1^4}{\Lambda_1^4 + (q_t^2-M^2_{D^{(*)}})^2} \right)^2,
\cr
F_{s,\,u}^C &=
\left( \frac{\Lambda_2^4}{\Lambda_2^4 + (q_s^2-M_N^2)^2} \right)
\left( \frac{\Lambda_2^4}{\Lambda_2^4 + (q_u^2-M^2_{\Lambda_c})^2} \right).
\nonumber \\
\label{FormFactor-1}
\end{align}

Next, the invariant amplitude $V_{\bar D^0 \Lambda_c^+ \to J/\psi p}$ 
can be written as
\begin{align}
V_{\bar D^0 \Lambda_c^+ \to J/\psi p} &=
\mathcal{M}_{V,t'}^D \, F_{t'}^D + \mathcal{M}_{V,t'}^{D^*} \, F_{t'}^{D^*}
\cr
& + (\mathcal{M}_{V,s'}^p + \mathcal{M}_{V,u'}^{\Lambda_c}) F_{s',u'}^C.
\label{V-Ampl-1}
\end{align}
in analogy with Eq.~(\ref{B-Ampl-1}).

The $D^0$- and $D^{*0}$-exchange amplitudes in the $t$ channel [Fig.~\ref{FIG02}(d)],
together with the proton- and $\Lambda_c^+$-exchange amplitudes in the $s$ and $u$
channels [Figs.~\ref{FIG02}(e,f)], are given by
\begin{align}
\mathcal{M}^D_{V,t'} &=
i \frac{g_{\psi D D} g_{D N \Lambda_c}}{q_t^{'2} - M_D^2}
\varepsilon^{*\mu} (q' - q'_t)_\mu
{\bar u}_N \gamma_5 u_{\Lambda_c},
\cr
\mathcal{M}^{D^*}_{V,t'} &=
\frac{g_{\psi D D^*} g_{D^* N \Lambda_c}}{q_t^{'2} - M_{D^*}^2}
\epsilon^{\mu\nu\alpha\beta}
{\bar u}_N \gamma_\mu q_{t \nu}' k_\alpha \varepsilon_\beta^* u_{\Lambda_c},
\cr
\mathcal{M}^p_{V,s'} &=
i \frac{g_{\psi NN} g_{D N \Lambda_c}}{q_s^{'2}-M_N^2} {\bar u}_N \slashed{\varepsilon}^*
(\slashed{q}_s' + M_N) \gamma_5 u_{\Lambda_c},
\cr
\mathcal{M}^{\Lambda_c}_{V,u'} &=
- i \frac{g_{\psi \Lambda_c \Lambda_c} g_{D N \Lambda_c}}{q_u^{'2}-M_{\Lambda_c}^2}
{\bar u}_N \gamma_5 (\slashed{q}_u' + M_{\Lambda_c}) \slashed{\varepsilon}^* u_{\Lambda_c},
\label{V-Ampl-2}
\end{align}
where $\varepsilon_\mu$ is the polarization vector of the outgoing $J/\psi$ meson.
The exchanged four-momenta are defined by
\begin{align}
q_t' = k - q',\,\,\, q_s' = k + p_f,\,\,\, q_u' = p_f - q'.
\label{FourMomenta-2}
\end{align}

For consistency, we adopt the same form for the form factors in
Eq.~(\ref{V-Ampl-1}) as in Eq.~(\ref{FormFactor-1}):
\begin{align}
F_{t'}^{D^{(*)}} &=
\left( \frac{\Lambda_1^4}{\Lambda_1^4 + (q_t^{\prime 2}-M^2_{D^{(*)}})^2}\right)^2,
\cr
F_{s',\,u'}^C &=
\left( \frac{\Lambda_2^4}{\Lambda_2^4 + (q_s^{\prime 2}-M_N^2)^2} \right)
\left( \frac{\Lambda_2^4}{\Lambda_2^4 + (q_u^{\prime 2}-M^2_{\Lambda_c})^2} \right).
\nonumber \\
\label{FormFactor-2}
\end{align}

\subsubsection{$\bar D^{*0} \Lambda_c^+$ intermediate state}

We next consider the invariant amplitude $B_{\gamma p \to \bar D^{*0} \Lambda_c^+}$.
For the $t$-channel diagram shown in Fig.~\ref{FIG02}(a), the amplitudes corresponding
to both $D^0$- and $D^{*0}$-exchanges are individually gauge invariant and are
included in this work.
By contrast, the $s$- and $u$-channel contributions in Figs.~\ref{FIG02}(b,\,c), as
in the $\bar D^0 \Lambda_c^+$ case, are not individually gauge invariant, although
their sum is.
Accordingly, the $\gamma p \to \bar D^{*0} \Lambda_c^+$ invariant amplitude is
constructed in a gauge-invariant form as
\begin{align}
B_{\gamma p \to \bar D^{*0} \Lambda_c^+} &=
\tilde{\mathcal M}_{B,t}^D \, F_t^D + \tilde{\mathcal M}_{B,t}^{D^*} \, F_t^{D^*}
\cr
& + (\tilde{\mathcal M}_{B,s}^p + \tilde{\mathcal M}_{B,u}^{\Lambda_c}) F_{s,\,u}^C.
\label{B-Ampl-3}
\end{align}
We use the same form factors as in Eq.~(\ref{FormFactor-1}), but with different
cutoff masses, $\Lambda_1 \to \Lambda_3,\,\Lambda_2 \to \Lambda_4$.
The relevant amplitudes are given by
\begin{align}
\tilde{\mathcal M}^{D^0}_{B,t} &=
i \frac{g_{\gamma D D^*} g_{D N \Lambda_c}}{q_t^2 - M_D^2}
\epsilon^{\mu\nu\alpha\beta} q_\mu q_\nu' \epsilon_\alpha \epsilon_\beta'^*
\bar u_{\Lambda_c} \gamma_5 u_N,
\cr
\tilde{\mathcal M}^{D^{*0}}_{B,t} &=
- \frac{g_{\gamma D^* D^*} g_{D^* N \Lambda_c}}{q_t^2 - M_{D^*}^2}
\bar u_{\Lambda_c} \left[ \gamma_\mu -\frac{1}{M_{D^*}^2} \slashed{q}_t q_{t\mu} \right]
u_N
\cr
& \times (q^\nu \epsilon^\mu -q^\mu \epsilon^\nu) \epsilon_\nu'^*,
\cr
\tilde{\mathcal M}^p_{B,s} &=
\frac{e g_{D^* N \Lambda_c}}{q_s^2-M_N^2}
\bar u_{\Lambda_c} \slashed{\epsilon}'^* (\slashed{q}_s + M_N)
\cr
& \times
\left[ \slashed{\epsilon} - \frac{\kappa_p}{4M_N}
(\slashed{\epsilon} \slashed{q} -\slashed{q} \slashed{\epsilon})
\right] u_N,
\cr
\mathcal{\tilde M}^{\Lambda_c}_{B,u} &=
\frac{e g_{D^* N \Lambda_c}}{q_u^2-M_{\Lambda_c}^2} \bar u_{\Lambda_c}
\left[ \slashed{\epsilon} - \frac{\kappa_{\Lambda_c}}{4M_{\Lambda_c}}
(\slashed{\epsilon} \slashed{q} -\slashed{q} \slashed{\epsilon}) \right]
\cr
& \times
(\slashed{q}_u + M_{\Lambda_c})
\slashed{\epsilon}'^* u_N,
\label{B-Ampl-4}
\end{align}
where the exchanged four-momenta are defined in Eq.~(\ref{FourMomenta-1}).
$\epsilon_\mu$ and $\epsilon_\nu'$ denote the polarization vectors of the incoming
photon and outgoing $D^*$ meson, respectively.

\begin{figure}[b]
\centering
\includegraphics[width=7.0cm]{FIG03.pdf}
\caption{The total cross section for $\gamma p \to J/\psi p$ as a function of the
laboratory photon energy $E_\gamma$. The dotted, dashed, and solid curves denote the
Pomeron contribution, the rescattering contribution, and their sum, respectively.
The results are compared with the SLAC~\cite{Camerini:1975cy},
Fermilab~\cite{Binkley:1981kv,E687:1993hlm}, H1~\cite{H1:1996kyo,H1:2000kis},
ZEUS~\cite{ZEUS:2002wfj}, GlueX~\cite{GlueX:2019mkq,GlueX:2023pev}, and
CLAS~\cite{CLAS:2026lls} data from threshold up to $E_\gamma \simeq 10^4$ GeV.}
\label{FIG03}
\end{figure}
\begin{figure*}[ht]
\centering
\includegraphics[width=7.0cm]{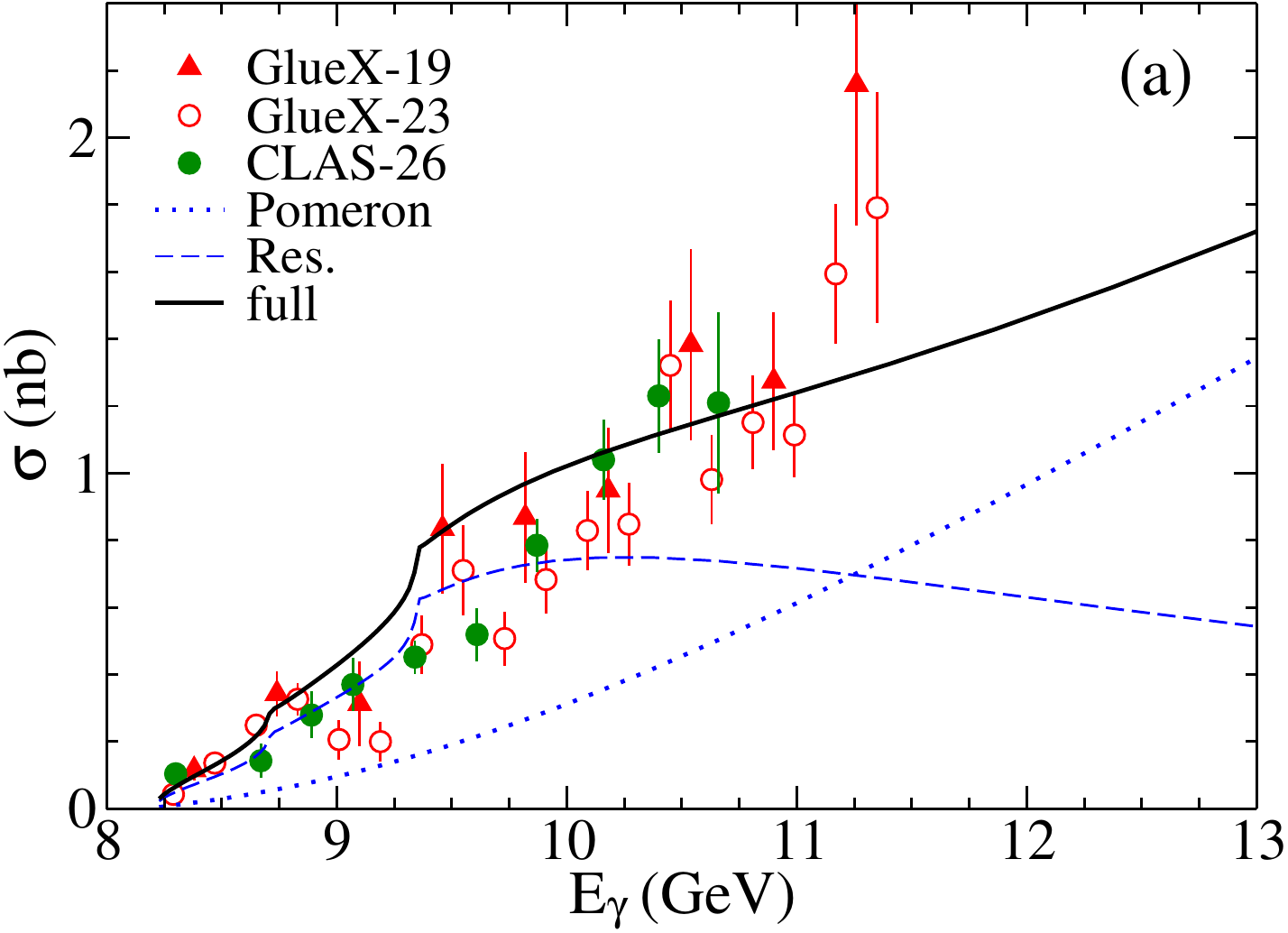} \,\,\,
\includegraphics[width=7.2cm]{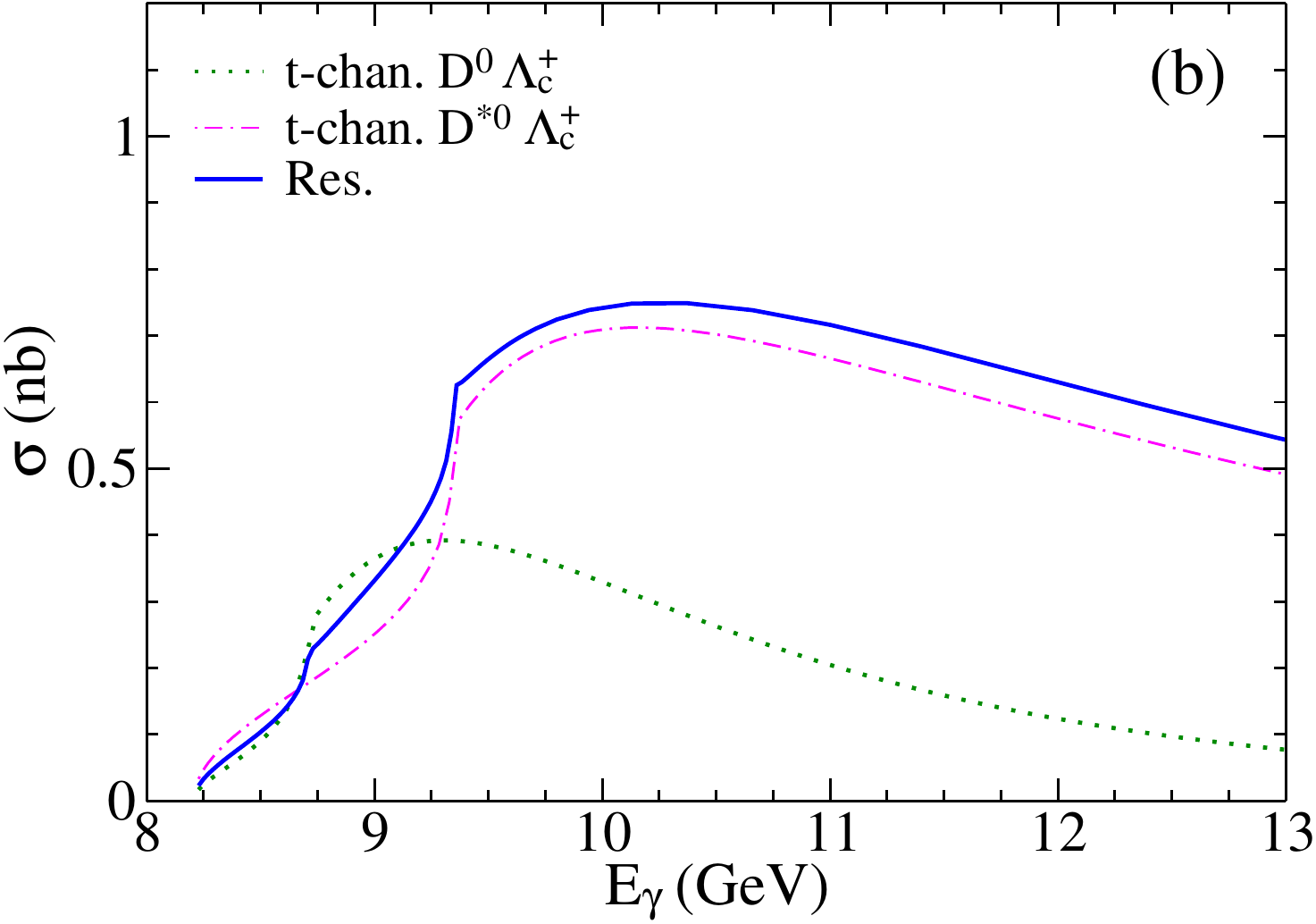}
\caption{{(a) The total cross section for $\gamma p \to J/\psi p$ as a function of the
laboratory photon energy $E_\gamma$, compared with the GlueX~\cite{GlueX:2019mkq,
GlueX:2023pev} and CLAS~\cite{CLAS:2026lls} data in the near-threshold region.
The blue dotted, dashed, and black solid curves represent the Pomeron contribution,
the rescattering contribution, and the full result, respectively.
(b) Decomposition of the rescattering contribution. The green dotted and magenta
dot-dashed curves represent the $t$-channel contributions from the
$\bar D^0 \Lambda_c^+$ and $\bar D^{*0} \Lambda_c^+$ intermediate states,
respectively, while the blue solid curve denotes the full rescattering result.}}
\label{FIG04}
\end{figure*}

Similarly, the invariant amplitude $V_{\bar D^{*0} \Lambda_c^+ \to J/\psi p}$ is constructed
as
\begin{align}
V_{\bar D^{*0} \Lambda_c^+ \to J/\psi p} &=
\tilde{\mathcal M}_{V,t'}^D \, F_{t'}^D + \tilde{\mathcal M}_{V,\,t'}^{D^*} \, F_{t'}^{D^*}
\cr
& + (\tilde{\mathcal M}_{V,\,s'}^p + \tilde{\mathcal M}_{V,\,u'}^{\Lambda_c}) F_{s',\,u'}^C.
\label{V-Ampl-3}
\end{align}
We use the same form factors as in Eq.~(\ref{FormFactor-2}), with the replacements
$\Lambda_1 \to \Lambda_3,\,\Lambda_2 \to \Lambda_4$.
The amplitudes corresponding to Figs.~\ref{FIG02}(d,\,e,\,f) are given by
\begin{align}
\tilde{\mathcal M}^{D^0}_{V,t} &=
- i \frac{g_{\psi D^* D} g_{D N \Lambda_c}}{q_t^{\prime 2} - M_D^2}
\epsilon^{\mu\nu\alpha\beta} q_\mu' k_\nu \epsilon'_\alpha \varepsilon_\beta^*
{\bar u}_N \gamma_5 u_{\Lambda_c},
\cr
\tilde{\mathcal M}^{D^{*0}}_{V,t} &=
\frac{g_{\psi D^* D^*} g_{D^* N \Lambda_c}}{q_t^{\prime 2} - M_{D^*}^2}
\bar u_N \left[ \gamma_\mu -\frac{1}{M_{D^*}^2} \slashed{q}'_t q_{t\mu}' \right]
u_{\Lambda_c}
\cr
& \times
[ (\varepsilon^* \cdot \epsilon') q'^\mu - (q_t' \cdot \epsilon') \varepsilon^{*\mu}
- \varepsilon^* \cdot (q' - q_t') \epsilon'^\mu],
\cr
\tilde{\mathcal M}^p_{V,s} &=
\frac{g_{\psi NN} g_{D^* N \Lambda_c}}{q_s^{\prime 2}-M_N^2}
{\bar u}_N \slashed{\varepsilon}^* (\slashed{q}_s' + M_N)
\slashed{\epsilon}' u_{\Lambda_c},
\cr
\tilde{\mathcal M}^{\Lambda_c}_{V,u} &=
- \frac{g_{\psi \Lambda_c \Lambda_c} g_{D^* N \Lambda_c}}{q_u^{\prime 2}-M_{\Lambda_c}^2}
{\bar u}_N \slashed{\epsilon}' (\slashed{q}_u' + M_{\Lambda_c})
\slashed{\varepsilon}^*  u_{\Lambda_c},
\label{V-Ampl-4}
\end{align}
where the exchanged four-momenta are defined in Eq.~(\ref{FourMomenta-2}), and
$\epsilon'_\mu$ and $\varepsilon_\nu$ denote the polarization vectors of the
incoming $D^*$ meson and outgoing $J/\psi$ meson, respectively.

\section{Numerical Results}
\label{Sec:III}

We now present and discuss our numerical results.
The cutoff masses in Eqs.~(\ref{FormFactor-1}) and (\ref{FormFactor-2}) are
determined by fitting to the near-threshold $\gamma p \to J/\psi p$ data~\cite{
GlueX:2019mkq,GlueX:2023pev,Duran:2022xag,007:2026dow,CLAS:2026lls} and are fixed as
\begin{align}
&\Lambda_1 = 1.58, \,\,\, \Lambda_2 = 2.35,
\cr
&\Lambda_3 = 1.18, \,\,\, \Lambda_4 = 2.25,
\end{align}
in units of GeV.

Figure~\ref{FIG03} shows the total cross section for $\gamma p \to J/\psi p$ as a
function of the laboratory (Lab) photon energy $E_\gamma$, calculated in a model
including Pomeron exchange and hadronic rescattering through the intermediate
$\bar D^0 \Lambda_c^+$ and $\bar D^{*0} \Lambda_c^+$ states, from threshold up to
$E_\gamma \simeq 10^4$ GeV.
The dotted, dashed, and solid curves represent the Pomeron contribution, the
rescattering contribution, and the full result, respectively.
At high energies, the cross section is dominated by Pomeron exchange within the DL
model~\cite{Donnachie:1984xq,Donnachie:1985iz,Donnachie:1987pu,Donnachie:1998gm,
Donnachie:1999qv} and is consistent with the available data~\cite{Binkley:1981kv,
E687:1993hlm,ZEUS:2002wfj,H1:1996kyo,H1:2000kis}.
Although the rescattering contribution becomes negligible at high energies, it plays
an important role near threshold, where it substantially modifies the energy
dependence of the cross section.
The convergence of the integration for the rescattering amplitude [see
Eq.~(\ref{eq:ResAmpl-1})] has been verified.

To illustrate the near-threshold effects of the rescattering mechanism,
Fig.~\ref{FIG04} presents the total cross section in the range 8 $\leqslant$
$E_\gamma$ $\leqslant$ 13 GeV.
As shown in Fig.~\ref{FIG04}(a), Pomeron exchange (blue dotted) alone underestimates
the JLab data~\cite{GlueX:2019mkq,GlueX:2023pev,CLAS:2026lls}, indicating that an
additional near-threshold mechanism is needed.
The meson--baryon rescattering contribution (blue dashed) generates pronounced
cusp-like structures near the $\bar D^0 \Lambda_c^+$ and $\bar D^{*0} \Lambda_c^+$
thresholds, thereby improving the description of the GlueX data~\cite{GlueX:2019mkq,
GlueX:2023pev}.
At the same time, the recent CLAS data~\cite{CLAS:2026lls} do not favor such
pronounced cusp structures and instead exhibit a smoother energy dependence, a trend
that is also compatible with the recent $J/\psi$-007 data~\cite{007:2026dow}.
Nevertheless, Fig.~\ref{FIG04}(a) clearly demonstrates that rescattering effects are
essential for understanding the near-threshold behavior of the cross section.

\begin{figure*}
\centering
\includegraphics[width=16.0cm]{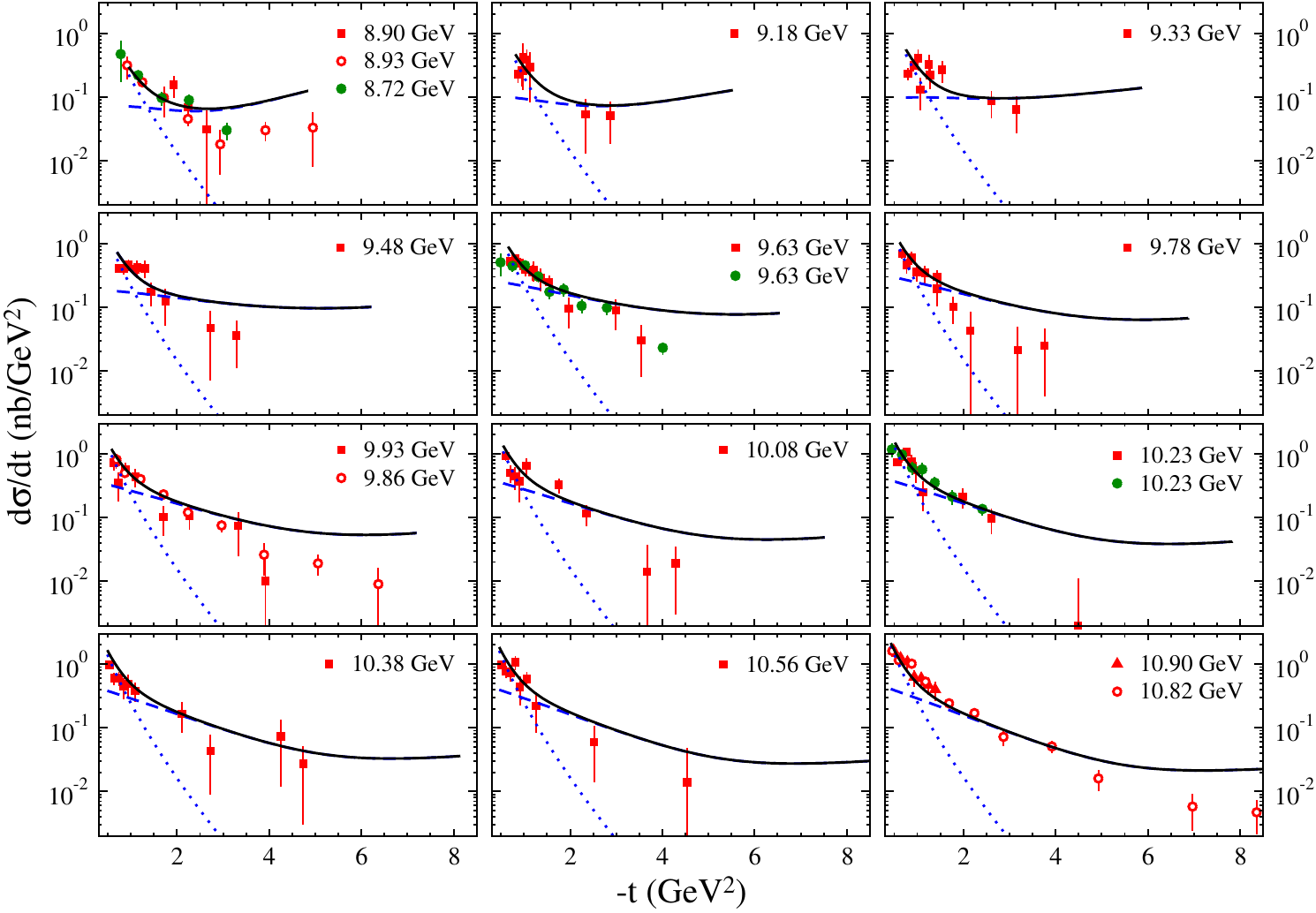}
\caption{Differential cross sections $d\sigma/dt$ as functions of $-t$ at various
photon energies, compared with the $J/\psi$-007~\cite{Duran:2022xag} (squares),
GlueX-23~\cite{GlueX:2023pev} (open circles), CLAS~\cite{CLAS:2026lls} (filled
circles), and GlueX-19~\cite{GlueX:2019mkq} (triangles) data.
The curve notations are the same as those in Fig.~\ref{FIG04}(a).}
\label{FIG05}
\end{figure*}
\begin{figure}[ht]
\centering
\includegraphics[width=7.0cm]{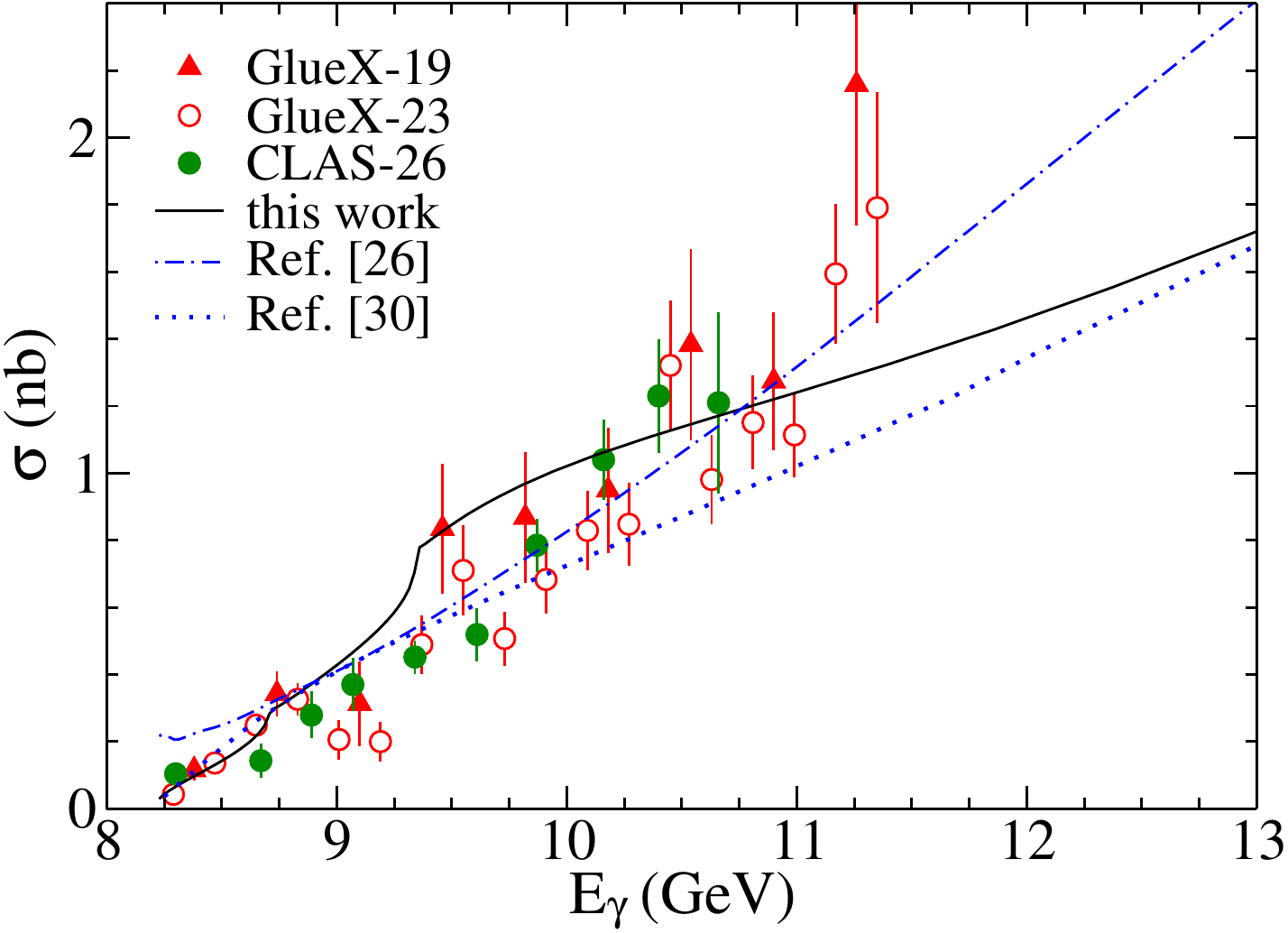}
\caption{The total cross section obtained in this work is compared with those of
Refs.~\cite{Sakinah:2024cza,Kim:2025oyo}.
Experimental data are taken from the GlueX~\cite{GlueX:2019mkq,GlueX:2023pev}
and CLAS~\cite{CLAS:2026lls} Collaborations.}
\label{FIG06}
\end{figure}
\begin{figure}[ht]
\centering
\includegraphics[width=5.5cm]{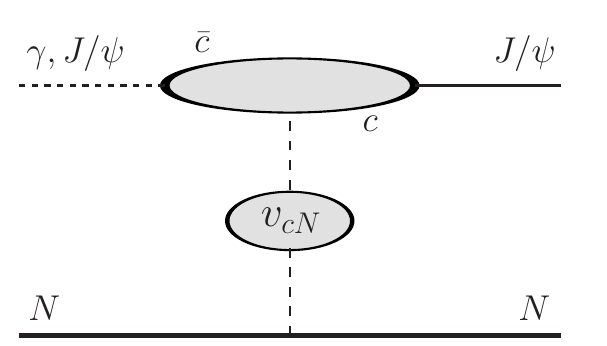}
\caption{Quark-antiquark loop mechanism for $\gamma, J/\psi + N \to J/\psi + N$ 
induced by a phenomenological charm-quark--nucleon potential $v_{cN}$.}
\label{FIG07}
\end{figure}

It is also interesting to note that another cusp-like structure 
appears around $E_\gamma\simeq$ 10.5 GeV in all JLab data~\cite{GlueX:2019mkq,
GlueX:2023pev,CLAS:2026lls}, corresponding to the $\bar D^{*0} \Sigma_c^{*+}$ threshold.
The description of the data may therefore be further improved by incorporating the
$\bar D^{*0} \Sigma_c^{*+}$ rescattering effect, although such an analysis is beyond
the scope of the present work.

\begin{figure*}[ht]
\centering
\includegraphics[width=16.0cm]{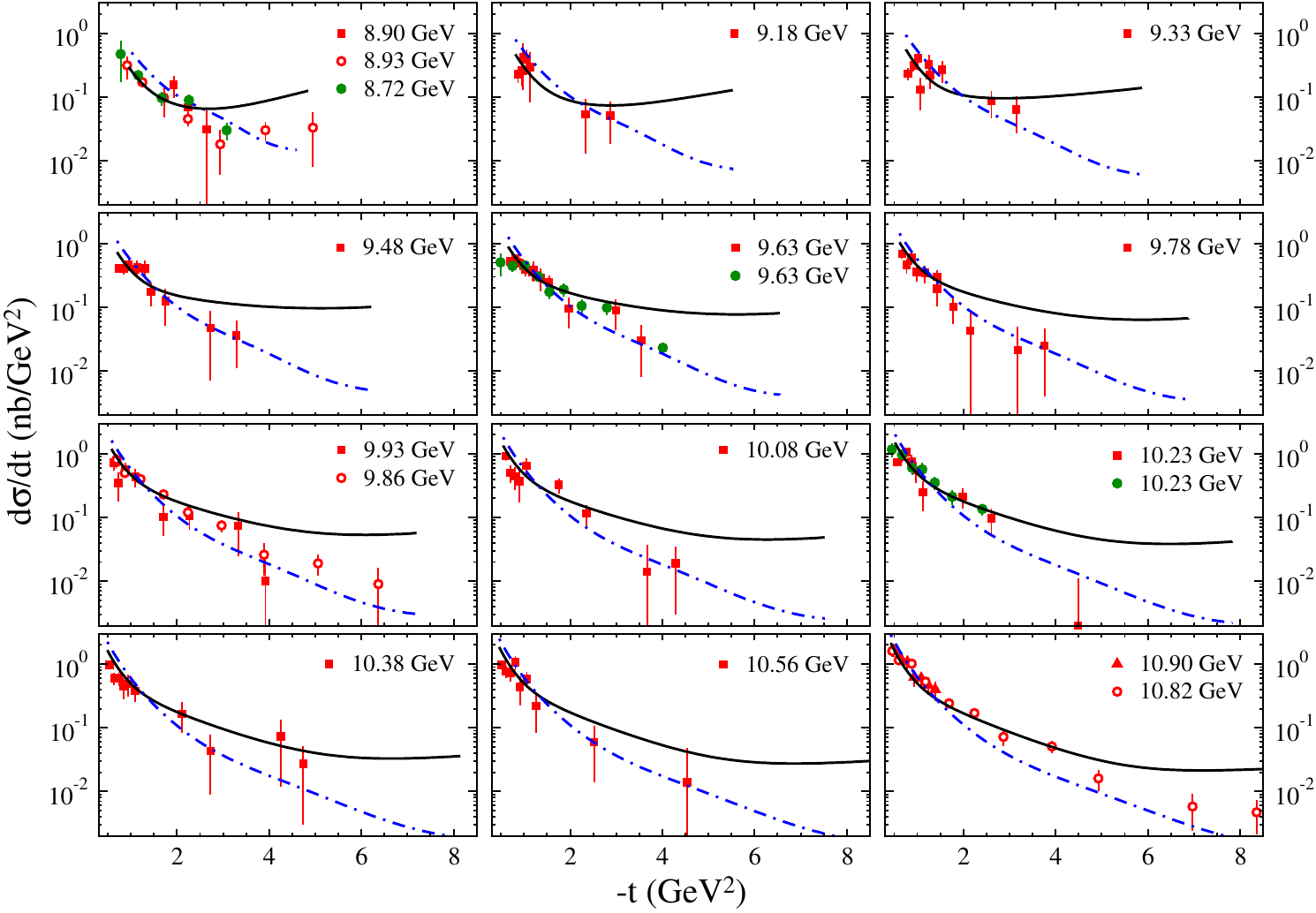}
\caption{Differential cross sections obtained in this work (black solid) are compared
with those of Ref.~\cite{Sakinah:2024cza} (blue dot-dashed).
Experimental data are taken from the $J/\psi$-007~\cite{Duran:2022xag} (squares),
GlueX-23~\cite{GlueX:2023pev} (open circles), CLAS~\cite{CLAS:2026lls} (filled
circles), and GlueX-19~\cite{GlueX:2019mkq} (triangles) Collaborations.}
\label{FIG08}
\end{figure*}
\begin{figure}[h]
\centering
\includegraphics[width=7.0cm]{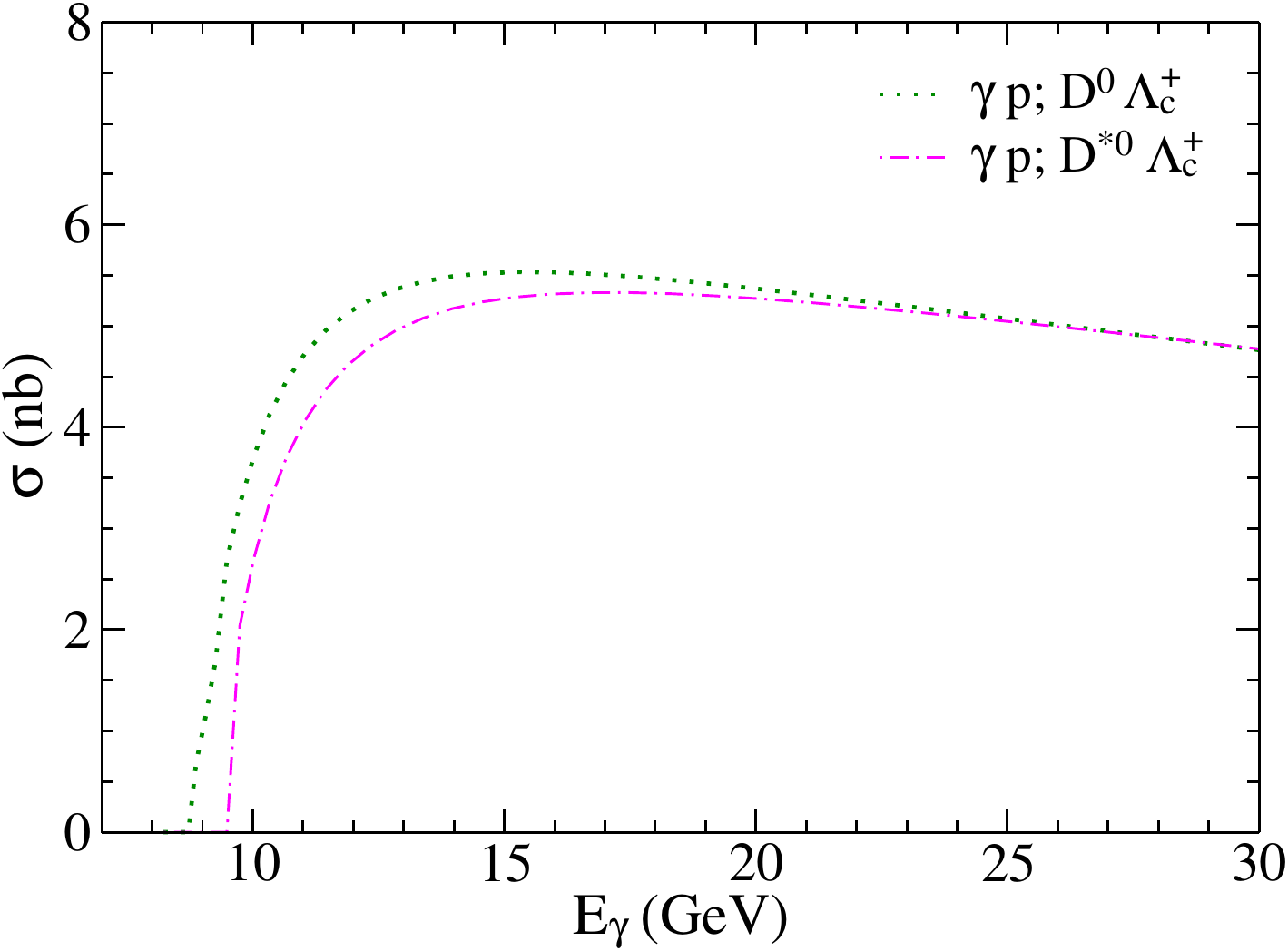}
\caption{Predicted total cross sections for the $\gamma p \to \bar D^0 \Lambda_c^+$
and $\gamma p \to \bar D^{*0} \Lambda_c^+$ reactions.}
\label{FIG09}
\end{figure}
The individual rescattering contributions are displayed in Fig.~\ref{FIG04}(b).
The blue solid curve represents the full rescattering result obtained by all $t$-,
$s$-, and $u$-channel diagrams in the $B$ and $V$ invariant amplitudes of
Fig.~\ref{FIG02}, and is consistent with the blue dashed curve in Fig.~\ref{FIG04}(a).
The green dotted and magenta dot-dashed curves show the contributions obtained by
retaining only the $t$-channel diagrams in Fig.~\ref{FIG02}(a,d) for the intermediate
$\bar D^0 \Lambda_c^+$ and $\bar D^{*0} \Lambda_c^+$ states, respectively.
These $t$-channel terms provide the dominant part of the rescattering contribution, 
whereas the $s$- and $u$-channel terms are suppressed by $3$--$7$ orders of magnitude
and decrease more rapidly with energy.
For clarity, the $s$- and $u$-channel contributions are not shown.

Figure~\ref{FIG05} presents the differential cross sections $d\sigma/dt$ at 12
different photon energies, compared with the $J/\psi$-007~\cite{Duran:2022xag}
(squares), GlueX-23~\cite{GlueX:2023pev} (open circles), CLAS~\cite{CLAS:2026lls}
(filled circles), and GlueX-19~\cite{GlueX:2019mkq} (triangles) data.
While the Pomeron contribution decreases rapidly with increasing $|t|$, the
rescattering contribution becomes increasingly important in the large-$|t|$ region.
As a result, the full calculation provides a much improved description of the JLab
data over the Pomeron contribution alone, especially at relatively large momentum
transfer.
Although the full results still tend to slightly overestimate some data points in the
backward region, they remain in reasonable agreement with the overall trend of the
data, particularly in the forward and intermediate regions.

Figure~\ref{FIG06} compares our results for the total cross section with those of
Refs.~\cite{Sakinah:2024cza,Kim:2025oyo}, both of which employ the same DL model for
the Pomeron-exchange process.
In Ref.~\cite{Sakinah:2024cza}, a dynamical model based on a phenomenological
charm-quark--nucleon ($c$-$N$) potential, $v_{cN}$, is additionally considered, as
illustrated in Fig.~\ref{FIG07}.
The potential is parametrized as
\begin{align}
v_{cN} (r) = \alpha \left( \frac{e^{-\mu_1 r}}{r}- \frac{e^{-\mu_2 r}}{r} \right),
\label{Potential-vcn}
\end{align}
where $\alpha$ = $-0.145$, $\mu_1$ = $0.3$ and $\mu_2$ = $1.5$~\cite{Sakinah:2024cza}.
The form of $v_{cN}(r)$ is chosen such that the resulting $V_{J/\psi N}(r)$ potential
reproduces the Yukawa-type behavior at large distances, consistent with lattice QCD
calculations~\cite{Kawanai:2010ev}.
It also provides a description of the JLab data including the $J/\psi N$ final-state
interaction.
The $VN$ potential is constructed by folding the charm-quark--nucleon interaction
$v_{cN}$ with the vector-meson wave function $\phi_V$ obtained from a constituent quark
model (CQM)~\cite{Segovia:2013wma}:
\begin{align}
V_{VN,VN}= \braket{\phi_V, N|\sum_{c}v_{cN}|\phi_V,N}.
\label{eq:v-vnvn}
\end{align}
In this approach, the commonly used VMD assumption is not employed.
By constrast, Ref.~\cite{Kim:2025oyo} incorporates a $t$-channel meson-exchange
mechanism, in which light-meson exchanges play an important role.
As shown in Fig.~\ref{FIG06}, the smooth energy dependence observed by the CLAS
Collaboration~\cite{CLAS:2026lls} is also supported by the results of Refs.~\cite{
Sakinah:2024cza} and \cite{Kim:2025oyo}.
The existence and interpretation of cusp-like structures in near-threshold $J/\psi$
photoproduction therefore remain open issues, which will be need to be clarified by
future experimental and theoretical studies.

Figure~\ref{FIG08} compares our results for the $t$-dependent differential cross
sections with those of Ref.~\cite{Sakinah:2024cza}.
The two theoretical results are, in general, consistent with the JLab data in the
forward region, corresponding to small momentum transfer
($-t \lesssim$ 2~GeV$^2$).
The results of Ref.~\cite{Sakinah:2024cza} exhibit a steeper falloff with increasing
$-t$ and remain in good agreement with the JLab data at relatively large momentum
transfer.

The size of the cusp structures is connected to the relevant open-charm reactions
$\gamma p \to \bar D^0 \Lambda_c^+$ and $\gamma p \to \bar D^{*0} \Lambda_c^+$.
Our predictions for their total cross sections are presented in Fig.~\ref{FIG09}.
The total cross sections increase with photon energy up to $E_\gamma \simeq$ 15 GeV,
reaching values of about $5$ nb, and then gradually decrease at higher energies.
Our predictions are smaller than those of Ref.~\cite{JPAC:2023qgg}, where total
cross sections of $\sigma \gtrsim$ 10 nb are  obtained, and are much smaller than
those of Ref.~\cite{Du:2019pij}, which predicts values on the order of $100$--$200$
nb.

\section{Summary and Conclusions}
\label{Sec:IV}

In this work, we investigated near-threshold $J/\psi$ photoproduction based on
the DL Pomeron-exchange mechanism supplemented by hadronic rescattering contributions
from the $\bar{D}^0\Lambda_c^+$ and $\bar{D}^{*0} \Lambda_c^+$ intermediate states.
The production and transition amplitudes were constructed by combining tree-level
subamplitudes in a gauge-invariant manner.
The rescattering amplitudes generate noticeable enhancements near $\bar{D}^0
\Lambda_c^+$ and $\bar{D}^{*0} \Lambda_c^+$ thresholds and improve the description of
the GlueX total cross-section data~\cite{GlueX:2019mkq,GlueX:2023pev}.
In addition, while the Pomeron contribution decreases rapidly with increasing $|t|$,
the open-charm rescattering contributions provide sizable effects at larger momentum
transfer, leading to a better description of the differential cross sections~\cite{
GlueX:2019mkq,Duran:2022xag,GlueX:2023pev,CLAS:2026lls}.
The dominant effect arises from the $t$-channel parts of the production and
transition subamplitudes, while the corresponding $s$- and $u$-channel contributions
are strongly suppressed.

Comparison with more recent JLab data suggests that the reaction mechanism for
near-threshold $J/\psi$ photoproduction is not yet fully understood.
In particular, a smoother energy dependence has been observed by the
$J/\psi$-007~\cite{007:2026dow} and CLAS~\cite{CLAS:2026lls} Collaborations.
Moreover, the structure around $E_\gamma \simeq$ 10.5 GeV~\cite{GlueX:2019mkq,
GlueX:2023pev,CLAS:2026lls} may reflect the influence of additional nearby
open-charm channels, such as $\bar{D}^{*0}\Sigma_c^{*+}$, which are not included in
the present analysis.
A more complete treatment of these additional channels will therefore be important for
achieving a deeper understanding of the threshold region.
If confirmed, such cusp-like structures would provide evidence for strong
coupled-channel dynamics and hadronic rescattering effects in near-threshold
$J/\psi$ photoproduction.
This would also make the extraction of gluonic properties of the nucleon, including
possible information related to the trace anomaly, more model dependent.

Similar cusp-like structures have also been discussed in the strange sector.
In $\phi N$ photoproduction~\cite{Kim:2021adl}, for example, a pronounced cusp appears
near $W \simeq 2.2$ GeV only at very forward angles ($\cos\theta > 0.9$)~\cite{
Dey:2014tfa}, suggesting an important role for intermediate strange-meson--baryon
loops, such as the $K^*\Sigma(1385)$, $K\Lambda(1670)$, and $K\Sigma(1660)$
channels.
A similar forward-angle enhancement has been reported in $K^+ \Sigma^0$
photoproduction near $W \simeq 1.9$ GeV~\cite{Jude:2020byj}.
These observations may indicate rescattering effects near the open- and
hidden-strangeness thresholds, in connection with hadronic bound states such as
$K \Lambda(1405)$, $K \Sigma(1385)$, and $\phi N$.
In $K^0 \Sigma^+$ photoproduction, on the other hand, the cross section drops
sharply near the $K^* \Lambda$ and $K^* \Sigma$ thresholds~\cite{
CBELSATAPS:2011gly}, although it remains unclear whether this behavior should be
interpreted as a cusp.
An alternative explanation has been proposed in terms of a $K^* \Sigma$ molecular
bound state in a coupled-channel approach~\cite{Nam:2021ayk,Khemchandani:2011et},
possibly related to the light pentaquark candidate $P_s^+(2071,\,3/2^-)$.
Such a state may be viewed as a light-flavor partner of the heavy pentaquark
$P_{c \bar c}(4457)$, although its spin-parity has not yet been determined.

These observations suggest that future measurements of open-charm photoproduction
reactions, such as $\gamma N \to \bar D^{(*)}\Lambda_c$ and related channels involving
$\Sigma_c$ baryons,
could provide useful information on the possible role of heavy-pentaquark states.
In particular, our model predicts that the total cross sections for $\gamma p \to
\bar{D}^{0}\Lambda_c^+$ and $\gamma p \to \bar{D}^{*0}\Lambda_c^+$ are of the order of
$5$ nb.
Measurements of these channels would provide a direct test of the rescattering
scenario explored in this work.
Extending the present model to the strangeness sector discussed above is also of
considerable interest, and work in this direction is currently underway.

\section*{Acknowledgments}

The authors are grateful to T.-S. H. Lee for fruitful discussions. 
This work was supported by the National Research Foundation of Korea (NRF) grant
funded by the Korea govenment under Grants No. RS-2026-25475296 (S.\,S., H.M.\,C)
and RS-2021-NR060129 (S.-H.\,K.).



\end{document}